\documentclass[twocolumn,showpacs,preprintnumbers,superscriptaddress,prl]{revtex4}

\usepackage{graphicx}
\usepackage{dcolumn}
\usepackage{bm}
\usepackage{color}
\usepackage{amssymb,amsmath,amsfonts}

\begin{document}

\title{Scaling and memory in the return intervals of energy dissipation rate in three-dimensional fully developed turbulence}

\author{Chuang Liu }
 \affiliation{School of Business, East China University of Science and Technology, Shanghai 200237, China} %
 \affiliation{Engineering Research Center of Process Systems Engineering (Ministry of Education), East China University of Science and Technology, Shanghai 200237, China} %
\author{Zhi-Qiang Jiang}
 \affiliation{School of Business, East China University of Science and Technology, Shanghai 200237, China} %
 \affiliation{School of Science, East China University of Science and Technology, Shanghai 200237, China} %
\author{Fei Ren}
 \affiliation{School of Business, East China University of Science and Technology, Shanghai 200237, China} %
 \affiliation{Engineering Research Center of Process Systems Engineering (Ministry of Education), East China University of Science and Technology, Shanghai 200237, China} %
 \affiliation{Research Center for Econophysics, East China University of Science and Technology, Shanghai 200237, China} %
\author{Wei-Xing Zhou }
 \email{wxzhou@ecust.edu.cn}
 \affiliation{School of Business, East China University of Science and Technology, Shanghai 200237, China} %
 \affiliation{Engineering Research Center of Process Systems Engineering (Ministry of Education), East China University of Science and Technology, Shanghai 200237, China} %
 \affiliation{School of Science, East China University of Science and Technology, Shanghai 200237, China} %
 \affiliation{Research Center for Econophysics, East China University of Science and Technology, Shanghai 200237, China} %
 \affiliation{Research Center on Fictitious Economics \& Data Science, Chinese Academy of Sciences, Beijing 100080, China} %

\date{\today}

\begin{abstract}
We study the statistical properties of return intervals $r$ between
successive energy dissipation rates above a certain threshold $Q$ in
three-dimensional fully developed turbulence. We find that the
distribution function $P_Q(r)$ scales with the mean return interval
$R_Q$ as $P_Q(r)=R_Q^{-1}f(r/R_Q)$ except for $r=1$, where the
scaling function $f(x)$ has two power-law regimes. The return
intervals are short-term and long-term correlated and possess
multifractal nature. The Hurst index of the return intervals decays
exponentially against $R_Q$, predicting that rare extreme events
with $R_Q\to\infty$ are also long-term correlated with the Hurst
index $H_\infty=0.639$.
\end{abstract}

\pacs{47.27.Jv, 47.53.+n, 05.45.Tp, 89.75.Da}

\maketitle

Extreme events are ubiquitous in nature and society and
understanding their dynamics is of crucial importance
\cite{Bunde-Kropp-Schellnhuber-2002,Kondratyev-Varotsos-Krapivin-2006}.
However, extreme events are usually rare, which makes it difficult
to investigate their occurrence properties. Recently, there is an
increasing interest in the study of return intervals or reoccurrence
times $r$ between successive events above (or below) some threshold
$Q$, aiming at unveiling the laws governing the occurrence of
extreme events by studying the statistics of the return intervals
for increasing threshold $Q$.

Recent studies show that the long-term correlation structure has
essential influence on the statistics of return intervals
\cite{Bogachev-Eichner-Bunde-PAG-2008}. For long-term power-law
correlated monofractal records with exponent $\gamma$, numerical
analysis illustrates that the distribution density function of the
return intervals follows a stretched exponential $P_Q(r)\sim
\exp\left[-b (r/R_Q)^{\gamma}\right]$ with the same exponent
$\gamma$ and the return intervals are also long-term correlated,
again with the same exponent $\gamma$, where $R_Q$ is the mean
return interval associated with threshold $Q$
\cite{Bunde-Eichner-Havlin-Kantelhardt-2003-PA,Bunde-Eichner-Kantelhardt-Havlin-2005-PRL,Altmann-Kantz-2005-PRE}.
Uncorrelated records with $\gamma=1$ is a special case, whose return
intervals are exponentially distributed. When $0<\gamma\leqslant1$,
theoretical analysis with certain approximation shows that the
distribution of return interval is Weibull: $P_Q(r) \sim
r^{\gamma-1} \exp(-c r^{\gamma})$
\cite{Olla-2007-PRE,Santhanam-Kantz-2008-PRE}. This seems consistent
with recent numerical results
\cite{Eichner-Kantelhardt-Bunde-Havlin-2007-PRE}.

For multifractal records in the presence or absence of linear
correlations, extensive simulations based on the multiplicative
random cascade (MRC) model \cite{Meneveau-Sreenivasan-1987-PRL} and
the multifractal random walk (MRW) model
\cite{Bacry-Delour-Muzy-2001-PRE} unveil that the return intervals
have a power-law decay in the distribution and are long-term
correlated governed by power laws whose exponents depend explicitly
on the threshold $Q$, and the conditional return intervals increase
as a power-law function of the previous return interval
\cite{Bogachev-Eichner-Bunde-2007-PRL,Bogachev-Eichner-Bunde-EPJST-2008,Bogachev-Bunde-2008-PRE}.
These results are of particular interest since a variety of time
series exhibit multifractal nature. For instance, the returns of
common stocks can be well modeled by the multifractal random walk
\cite{Muzy-Delour-Bacry-2000-EPJB,Bacry-Delour-Muzy-2001-PA}, and
the statistics of the associated return intervals are found to
comply with the numerical prediction
\cite{Bogachev-Eichner-Bunde-2007-PRL,Bogachev-Bunde-2008-PRE},
which can be used to significantly improve risk estimation
\cite{Bogachev-Eichner-Bunde-2007-PRL}.

Another important issue of multifractal records concerns the
possible scaling behavior of the return interval distributions
$P_Q(r)$ over different thresholds $Q$. Numerical simulations of MRC
and MRW time series find no evidence of such scaling
\cite{Bogachev-Eichner-Bunde-2007-PRL,Bogachev-Eichner-Bunde-EPJST-2008,Bogachev-Bunde-2008-PRE}.
However, empirical return interval analysis of financial volatility
gives miscellaneous results. Several studies reported that there is
a scaling law in the return interval distributions
\cite{Yamasaki-Muchnik-Havlin-Bunde-Stanley-2005-PNAS,Wang-Yamasaki-Havlin-Stanley-2006-PRE,Wang-Weber-Yamasaki-Havlin-Stanley-2007-EPJB,Jung-Wang-Havlin-Kaizoji-Moon-Stanley-2008-EPJB,Qiu-Guo-Chen-2008-PA},
while others argued that the cumulative distributions of return
intervals had systematic deviations from scaling and showed
multiscaling behaviors
\cite{Wang-Yamasaki-Havlin-Stanley-2008-PRE,Lee-Lee-Rikvold-2006-JKPS,Ren-Zhou-2008-EPL,Ren-Guo-Zhou-2009-PA}.

In this Letter, we perform return interval analysis of the energy
dissipation rate in three-dimensional fully developed turbulence
based on a high-Reynolds turbulence data set collected at the S1
ONERA wind tunnel by the Grenoble group from LEGI
\cite{Anselmet-Gagne-Hopfinger-Antonia-1984-JFM}. The size of the
velocity time series $\{v_i: i=1,2,\cdots,N\}$ is about $1.73 \times
10^7$. Using Taylor's frozen flow hypothesis which replaces a
spatial variation of the fluid velocity by a temporal variation
measured at a fixed location, the rate of kinetic energy dissipation
at position $i$ is $ \epsilon_i \sim \left[ \left(v_{i+1} - v_i
\right) / \delta_\ell \right] ^2$, where $\delta_\ell$ is the
resolution (translated in spatial scale) of the measurements. The
energy dissipation rate time series exhibit multifractal nature
\cite{Zhou-Sornette-2002-PD} and its Hurst index is $H =
1-\gamma/2=0.81$. Contrary to previous studies, we find that the
return interval distributions show two power-law regimes and
collapse onto a single curve for different thresholds $Q$. The
scaling phenomenon is also observed for conditional interval
distributions.

\begin{figure}[htb]
\centering
\includegraphics[width=7cm]{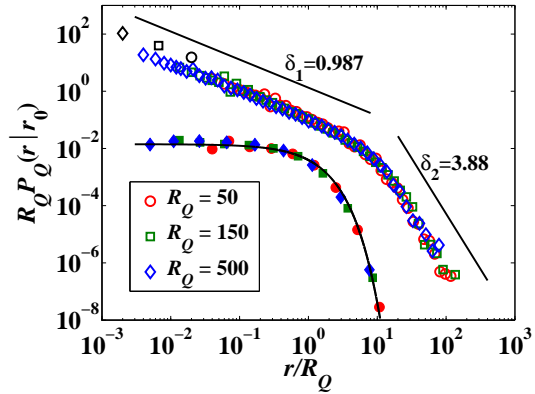}
\caption{\label{Fig:RI:PDF} (color online). Log-log plot of the
scaled distribution $R_QP_Q(r)$ as a function of $r/R_Q$ for three
different values of $R_Q$. The straight lines are the best fits to
power laws. The power-law exponents are $\delta_1=0.987\pm0.013$ and
$\delta_2=3.88 \pm 0.09$. For the shuffled data, the $R_QP_Q(r)$
curves collapse to a same exponential curve, as expected.}
\end{figure}

We have calculated the return interval time series $r_i$ for
different thresholds $Q$, which can be mapped nonlinearly to the
mean return intervals $R_Q$. Logarithmic binning is adopted to
construct the distribution density functions $P_Q(r)$. In order to
ensure that the bins cover the whole $r$-axis, we use the following
procedure. First, the interval $[1,\max(r_i)]$ is partitioned
logarithmically into $n-1$ subintervals whose edges are
$x_1<x_2<\cdots<x_n$. Then we obtain the sequence $y_i=[x_i]$, where
$[x_i]$ is a round function of $x_i$. We discard duplicate integers
in the $y_i$ sequence and obtain a new sequence $w_j$. The edge
sequence of the bins are determined by
$\{e_i\}=\{0.5,\{w_i+0.5\}\}$. For each bin $(e_i,e_{i+1}]$, the
empirical density function can be calculated by
\begin{equation}
 P_Q(r_i) = \frac{\#(e_i<r<e_{i+1})}{\#(r>0)}\frac{1}{e_{i+1}-e_i},
 \label{Eq:Pq:r}
\end{equation}
where $r_i=(e_i+e_{i+1})/2$ and $\#()$ is the number of return
intervals that satisfies the condition in the parenthesis. The
empirical distribution of the return intervals is depicted in
Fig.~\ref{Fig:RI:PDF} for $R_Q=50$, $150$ and $500$. We find that
the three distributions collapse onto a single curve except for
$r=1$, in remarkable contrast to the simulation results for MRC and
MRW time series. Two power-law regimes are observed
\begin{equation}
 R_QP_Q(r)\sim \left\{
 \begin{array}{lllll}
    A_1(r/R_Q)^{-\delta_1} && {\rm{if}}& 1<r<r_c\\
    A_2(r/R_Q)^{-\delta_2} && {\rm{if}}& r>r_c\\
 \end{array}
 \right.,
 \label{Eq:RI:PDF}
\end{equation}
where $A_1=0.107$ and $A_2=33.4$ are prefactors, the crossover
return interval $r_c\approx7R_Q$, and $\delta_1=0.987\pm0.013$ and
$\delta_2=3.88 \pm 0.09$. For the shuffled data, the $R_QP_Q(r)$
curves collapse to a single exponential curve.

In risk estimation, a quantity of great interest is the probability
$W_Q(\Delta{t},t)$ that an extreme event occurs after a short time
$\Delta{t}\ll t$ from now on, conditioned that the time elapsed $t$
after the occurrence of the previous extreme event
\cite{Bogachev-Eichner-Bunde-2007-PRL}:
\begin{equation}
 W_Q(\Delta{t}|t)=\frac{\int_t^{t+\Delta{t}}P_Q(t)dt}{\int_t^{\infty}P_Q(t)dt}.
 \label{Eq:Wq}
\end{equation}
When $t>r_c$, simple algebraic manipulation leads to
\begin{equation}
 W_Q(\Delta{t}|t)\simeq(\delta_2-1){\Delta{t}}/{t}.
 \label{Eq:Wq2}
\end{equation}
The probability $W_Q(\Delta{t}|t)$ is found to be proportional to
$\Delta{t}$ and inversely proportional to $t$. An intriguing feature
is that $W_Q(\Delta{t}|t)$ is independent of the threshold $Q$,
which is a direct consequence of the scaling behavior of $P_Q(r)$
shown in Fig.~\ref{Fig:RI:PDF}. When $t<r_c$, we obtain that
\begin{eqnarray}
 W_Q(\Delta{t}|t)&\approx&\frac{(\delta_1-1)\left(\frac{t}{R_Q}\right)^{-\delta_1}\frac{\Delta{t}}{R_Q}}
                       {\left(\frac{t}{R_Q}\right)^{1-\delta_1}
                       +\frac{A_2}{A_1}\frac{\delta_1-1}{\delta_2-1}\left(\frac{r_c}{R_Q}\right)^{1-\delta_2}
                       -\left(\frac{r_c}{R_Q}\right)^{1-\delta_1}
                       }\nonumber\\
                 &\approx& \frac{(\delta_1-1)\left(\frac{t}{R_Q}\right)^{-\delta_1}\frac{\Delta{t}}{R_Q}}
                       {\left(\frac{t}{R_Q}\right)^{1-\delta_1}-1.0308}.
 \label{Eq:Wq1}
\end{eqnarray}
We find that $W_Q(\Delta{t}|t)$ is proportional to $\Delta{t}$.
However, $W_Q(\Delta{t}|t)$ also depends on $R_Q$.

\begin{figure}[htb]
\centering
\includegraphics[width=7cm]{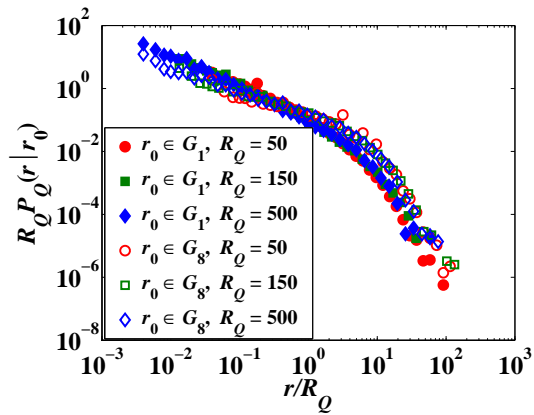}
\caption{\label{Fig:RI:CondPDF} (color online). Scaling and memory
in the conditional distributions. The scaled conditional
distribution $R_QP_Q(r|r_0)$ is plotted as a function of $r/R_Q$
with $r_{0}$ in $G_1$ and $G_8$ for three different values of
$R_Q$.}
\end{figure}

In order to test the memory effects of the return intervals, we
first investigate the conditional PDF $P_Q(r|r_0)$, which is the
distribution of return intervals immediately after $r_0$. To gain
better statistics, we study $P_Q(r|r_0)$ for a range of $r_0$ rather
than individual $r_0$ values. For each threshold $Q$ or $R_Q$, the
return intervals sequence are sorted in an increasing order and then
divided into eight groups $G_1,\cdots,G_8$ with approximately equal
size. An empirical conditional distribution is determined for each
$r_0$ group. Figure \ref{Fig:RI:CondPDF} shows $P_Q(r|r_0)$ for
$r_0\in G_1$ and $r_0\in G_8$. For each group, the three
distributions collapse onto a single curve, indicating evident
scaling behavior. The figure shows that the probability of finding
small (large) $r$ in $G_1$ is enhanced (decreased) compared with
$G_8$. This discrepancy in the two groups of distributions unveils
the memory effect that large (small) return intervals tend to follow
large (small) return intervals. This is true since the distributions
associated with different $G_i$ should not exhibit significant
discrepancy if there is no memory in the return intervals
\cite{Yamasaki-Muchnik-Havlin-Bunde-Stanley-2005-PNAS}.

\begin{figure}[htb]
\centering
\includegraphics[width=7cm]{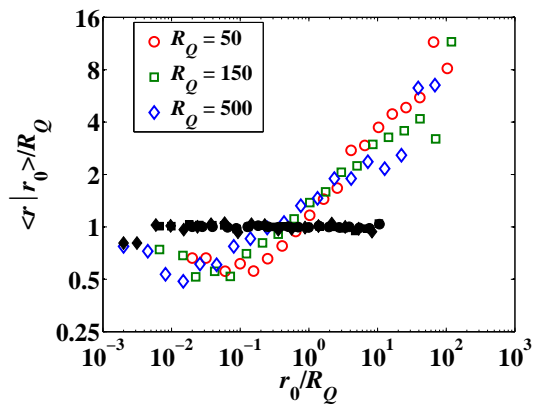}
\caption{\label{Fig:RI:Mean} (color online). Scaled mean conditional
return interval $\langle r|r_0\rangle/R_Q$ versus $r_0/R_Q$ of
return interval sequence in turbulence for various $R_Q$. The filled
symbols represent the results for the shuffled records without
memory. }
\end{figure}

The memory effect in the conditional distribution $P_Q(r|r_0)$ can
also be illustrated by the mean conditional return interval $\langle
r|r_0\rangle$. If there is no memory in the return intervals,
$\langle r|r_0\rangle$ does not depend on $r_0$ and is equal to
$R_Q$. Figure \ref{Fig:RI:Mean} plots the mean conditional return
interval $\langle r|r_0\rangle$ as a function of $r_0$ for $R_Q=50$,
150 and 500. It is shown that $\langle r|r_0\rangle$ is a power-law
function of $r_0$ when $r_0$ is larger than certain value and the
power-law exponent decreases with $R_Q$, which is consistent with
many simulational and empirical studies. Also shown in
Fig.~\ref{Fig:RI:Mean} is the mean conditional return interval of
the shuffled energy dissipation rate, which does not depend on
$r_0$.

We now study the long-term correlation in the return intervals using
the multifractal detrended fluctuation analysis (MFDFA), which is
able to extract long-term power-law correlation in non-stationary
time series
\cite{Peng-Buldyrev-Havlin-Simons-Stanley-Goldberger-1994-PRE,Kantelhardt-Bunde-Rego-Havlin-Bunde-2001-PA,Kantelhardt-Zschiegner-Bunde-Havlin-Bunde-Stanley-2002-PA}.
The MFDFA considers the cumulative time series $R_i=\sum_{i=1}^m
(r_i-\langle{r}\rangle)$, which is partitioned into $N_s$ disjoint
boxes with the same size $s$. In each box $k$, the local trend is
removed from the subseries by a polynomial function and the local
rms fluctuation $f_k(s)$ is determined. The overall detrended
fluctuation is calculated by
\begin{equation}
 F_q(s) = \left\{\frac{1}{N_s}\sum_{k=1}^{N_s}[f_k(s)]^q
 \right\}^{1/q}.
  \label{Eq:Fq}
\end{equation}
By varying the value of $s$, one can expect the detrended
fluctuation function $F_q(s)$ scales with the size $s$:
\begin{equation}
 F_q(s) \sim s^{h(q)},
  \label{Eq:scaling}
\end{equation}
where $h(q)$ is the generalized Hurst index. The return interval
series possesses multifractal nature if and only if $h(q)$ is a
nonlinear function of $q$. When $q = 2$, $h(2)$ is nothing but the
Hurst index $H$ and the MFDFA reduces to the DFA. The Hurst index
$H$ is related to the autocorrelation exponent $\gamma$ by $\gamma =
2-2H$.

\begin{figure}[htb]
\centering
\includegraphics[width=7cm]{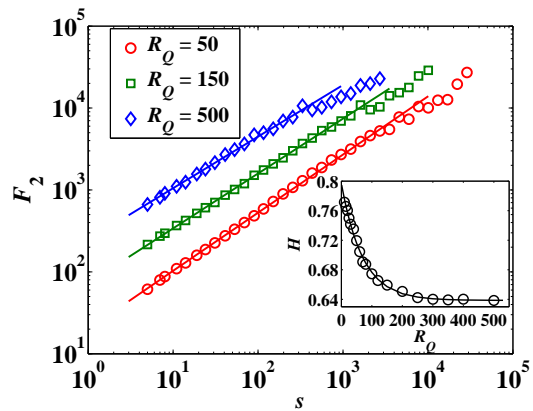}
\caption{\label{Fig:RI:DFA} (color online). Detrended fluctuation
analysis of the return interval time series for different $R_Q$. The
inset shows the exponential decay of the Hurst index $H$ against
$R_Q$.}
\end{figure}

We first investigate the linear long-term correlation property of
the return intervals using DFA. The dependence of the fluctuation
function $F_2$ is drawn in Fig.~\ref{Fig:RI:DFA} against $s$ for
$R_Q=50$, 150, and 500. In all cases, we find nice power-law
relation and the scaling range decreases with the increase of $R_Q$.
The inset shows the dependence of the Hurst index with respect to
$R_Q$, which has an exponential decay:
\begin{equation}
  H = H_{\infty} + b e^{-R_Q/R_c} = 0.639 + 0.158 e^{-R_Q/69.9},
  \label{Eq:H:Rq}
\end{equation}
where $R_c=69.9$ is the characteristic scale. For extreme events
with very large $Q$, $R_Q$ tends to infinity, and the Hurst index
can be predicted as $H=H_{\infty}=0.639$. This implies that the
return intervals of those extreme events also exhibit long-term
memory.

\begin{figure}[htb]
\centering
\includegraphics[width=4cm]{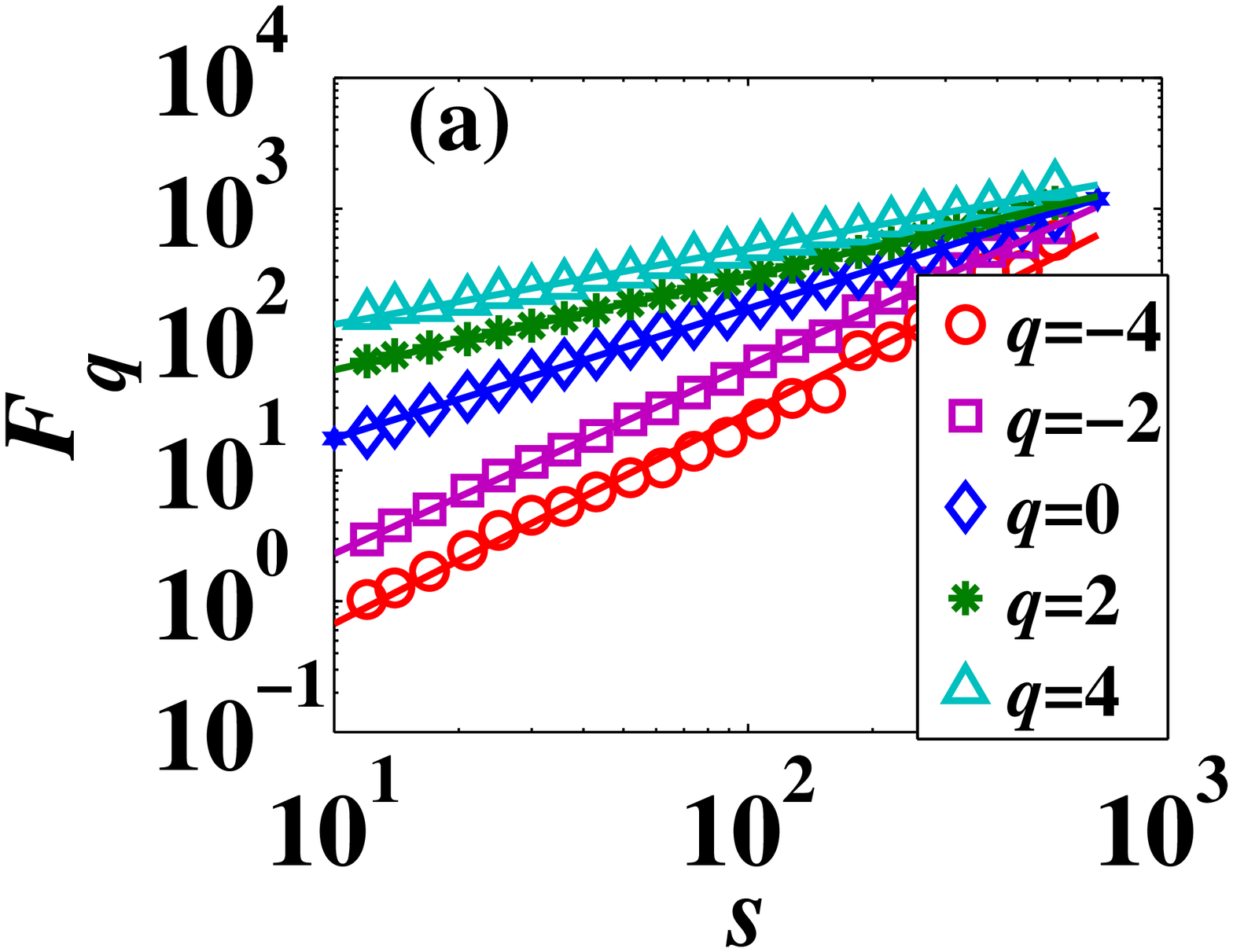}
\includegraphics[width=4cm]{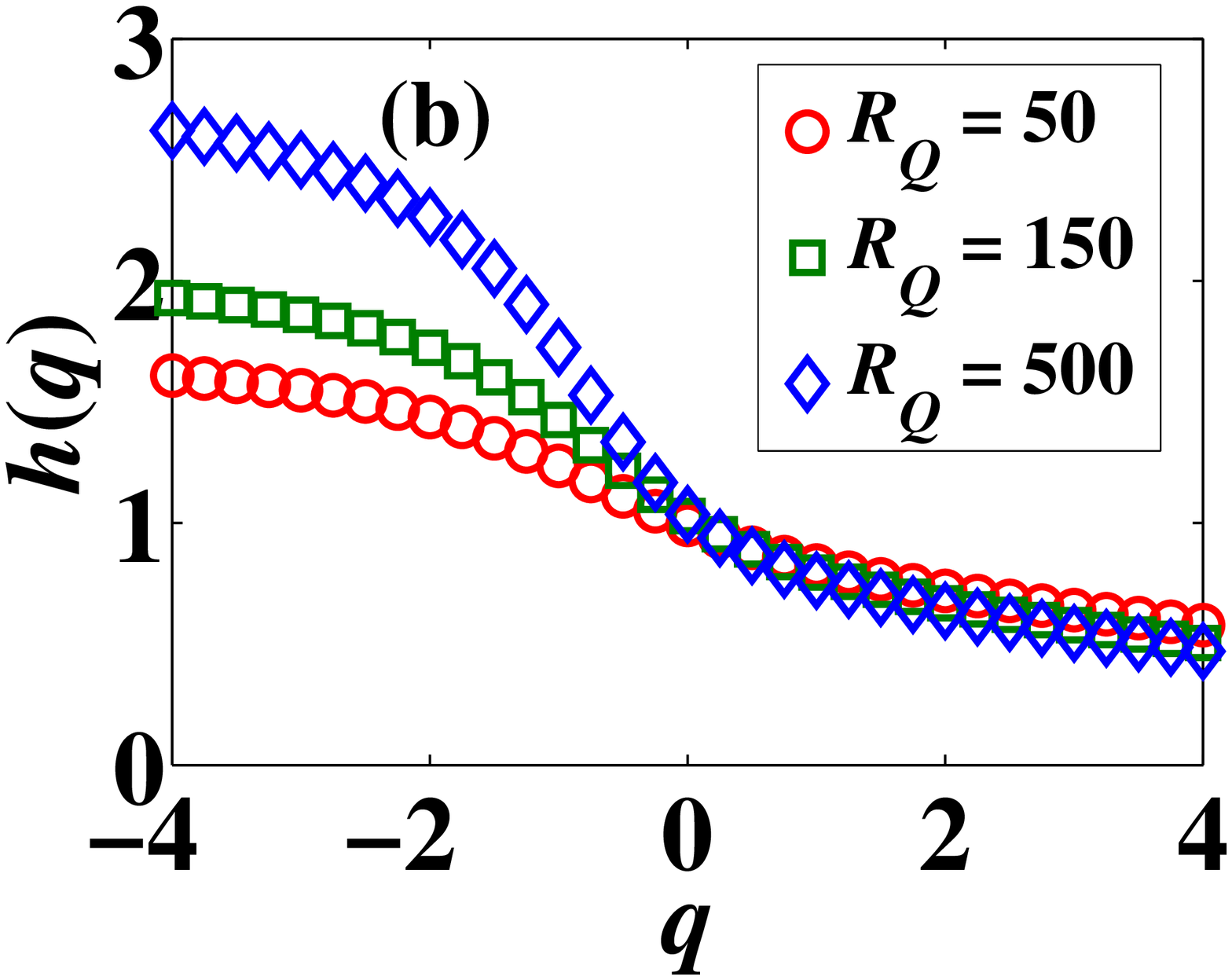}
\includegraphics[width=4cm]{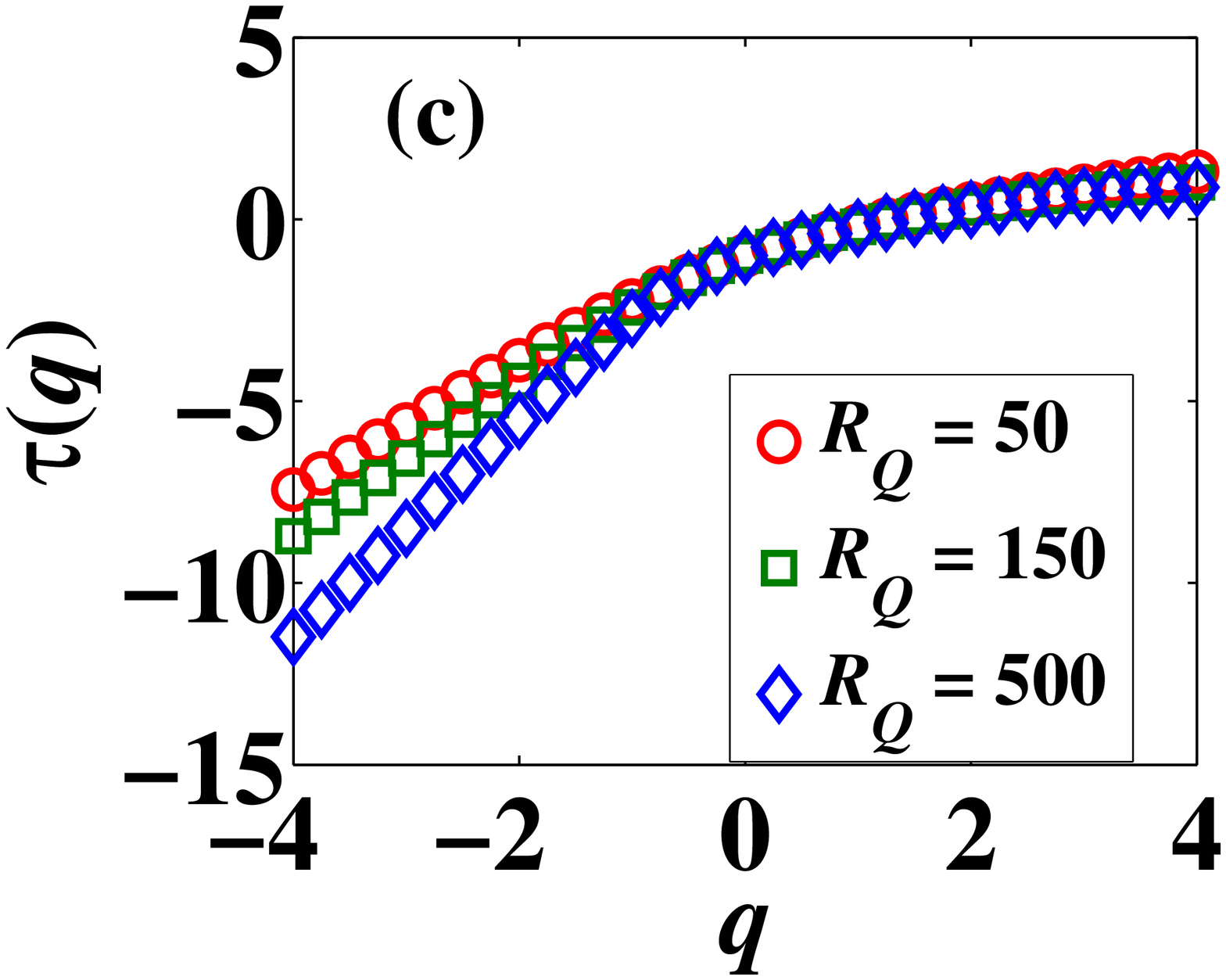}
\includegraphics[width=4cm]{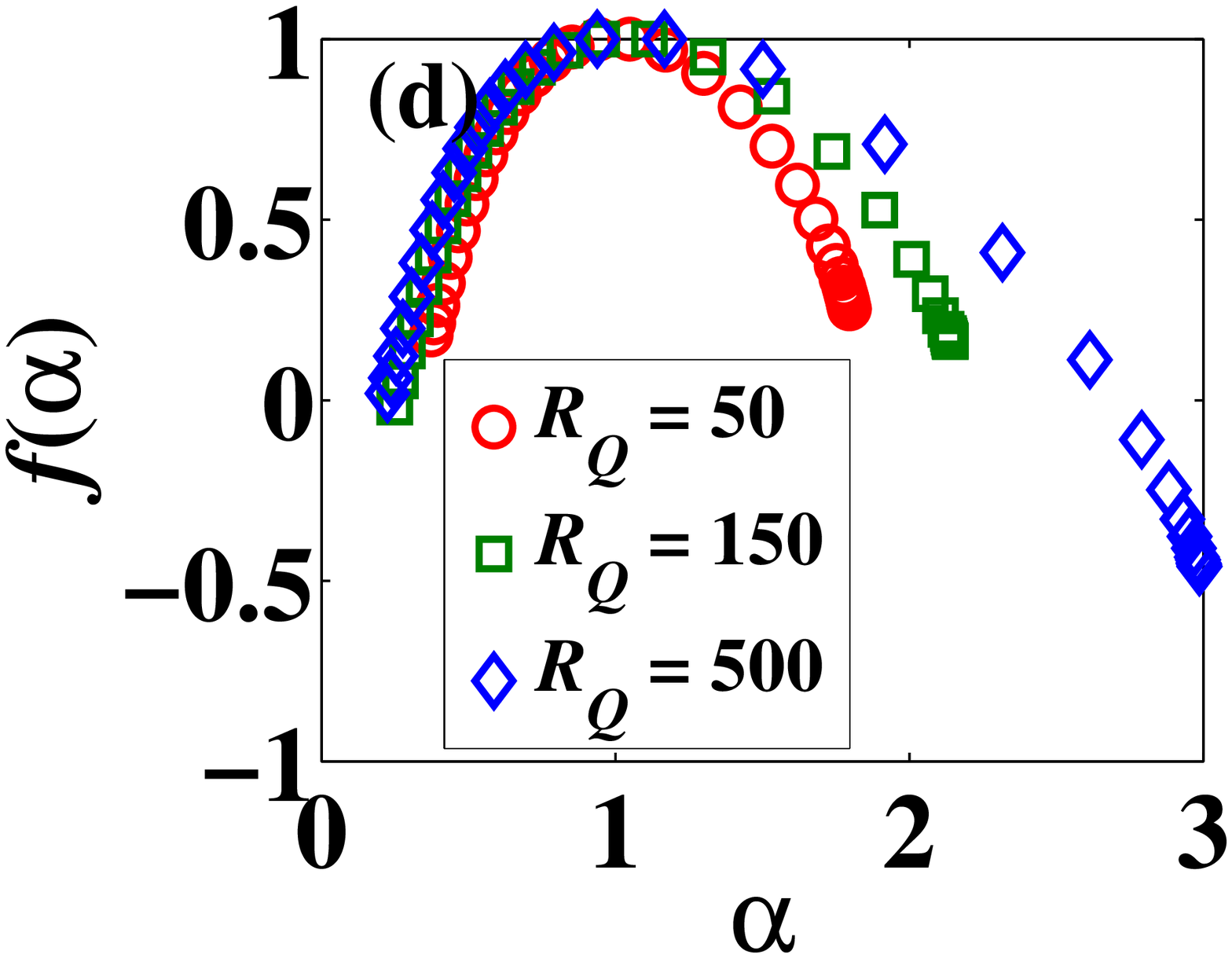}
\caption{\label{Fig:RI:MFDFA} (color online). Multifractal detrended
fluctuation analysis of the return interval time series for
different $R_Q$. (a) MFDFA fluctuation function for different $q$
when $R_Q=50$. (b) Generalized Hurst indexes $h(q)$. (c) Mass
scaling exponents $\tau(q)$. (d) Multifractal singularity spectra
$f(\alpha)$.}
\end{figure}

We also apply the MFDFA to investigate the multifractal nature of
the return intervals. Figure \ref{Fig:RI:MFDFA}(a) shows the
power-law dependence of the overall fluctuation $F_q(s)$ on the
scale $s$ for $R_Q=50$. When $R_Q$ increases, the $F_q(s)$ function
becomes more noisy, especially for negative $q$. The slopes of the
straight lines are the linear least-squares estimates of the
generalized Hurst indexes $h(q)$, which are drawn in
Fig.~\ref{Fig:RI:MFDFA}(b). The mass scaling exponent function
$\tau(q)=qh(q)-1$ and the multifractal spectrum $f(\alpha)$
calculated according to the Legendre transform of $\tau(q)$ are
illustrated respectively in Fig.~\ref{Fig:RI:MFDFA}(c) and
Fig.~\ref{Fig:RI:MFDFA}(d). The sound nonlinearity in $h(q)$ and
$\tau(q)$ is a hallmark of multifractality in the return intervals,
whose singularity strength increases with $R_Q$.

In summary, we have studied the statistical properties of return
intervals of the energy dissipation rate in three-dimensional fully
developed turbulence. The interval distribution is found to exhibit
scaling behavior across different $R_Q$ and two power-law regimes,
except for intervals $r=1$. We found that the conditional interval
distributions also collapse onto a single curve for same $r_0$, but
deviate for different $r_0$, and the mean conditional interval
increases as a power law of $r_0$, which indicates the presence of
short memory in the return intervals. The long-term memory and the
multifractal nature in the return intervals are confirmed by the DFA
and MFDFA. The Hurst index of the return intervals decays
exponentially against $R_Q$, which allows us to predict the
asymptotic Hurst index of return intervals between rare extreme
events as $H_\infty=0.639$. Our results signify discrepancy with the
numerical results using MRC and MRW models
\cite{Bogachev-Eichner-Bunde-2007-PRL}, implying that the energy
dissipation process can not be modeled using these models.

\begin{acknowledgements}
This work was partially supported by the National Basic Research
Program of China (2004CB217703), the Program for Changjiang Scholars
and Innovative Research Team in University (IRT0620), and the
Program for New Century Excellent Talents in University
(NCET-07-0288).
\end{acknowledgements}

\bibliography{E:/Papers/Bibliography}

\begin{thebibliography}{30}
\expandafter\ifx\csname natexlab\endcsname\relax\def\natexlab#1{#1}\fi
\expandafter\ifx\csname bibnamefont\endcsname\relax
  \def\bibnamefont#1{#1}\fi
\expandafter\ifx\csname bibfnamefont\endcsname\relax
  \def\bibfnamefont#1{#1}\fi
\expandafter\ifx\csname citenamefont\endcsname\relax
  \def\citenamefont#1{#1}\fi
\expandafter\ifx\csname url\endcsname\relax
  \def\url#1{\texttt{#1}}\fi
\expandafter\ifx\csname urlprefix\endcsname\relax\def\urlprefix{URL }\fi
\providecommand{\bibinfo}[2]{#2}
\providecommand{\eprint}[2][]{\url{#2}}

\bibitem[{\citenamefont{Bunde et~al.}(2005{\natexlab{a}})\citenamefont{Bunde,
  Kropp, and Schellnhuber}}]{Bunde-Kropp-Schellnhuber-2002}
\bibinfo{editor}{\bibfnamefont{A.}~\bibnamefont{Bunde}},
  \bibinfo{editor}{\bibfnamefont{J.}~\bibnamefont{Kropp}}, \bibnamefont{and}
  \bibinfo{editor}{\bibfnamefont{H.-J.} \bibnamefont{Schellnhuber}}, eds.,
  \emph{\bibinfo{title}{{The Science of Disasters}}}
  (\bibinfo{publisher}{Springer}, \bibinfo{address}{Berlin},
  \bibinfo{year}{2005}{\natexlab{a}}).

\bibitem[{\citenamefont{Kondratyev et~al.}(2006)\citenamefont{Kondratyev,
  Varotsos, and Krapivin}}]{Kondratyev-Varotsos-Krapivin-2006}
\bibinfo{author}{\bibfnamefont{K.~Y.} \bibnamefont{Kondratyev}},
  \bibinfo{author}{\bibfnamefont{C.~A.} \bibnamefont{Varotsos}},
  \bibnamefont{and} \bibinfo{author}{\bibfnamefont{V.~F.}
  \bibnamefont{Krapivin}}, \emph{\bibinfo{title}{{Natural Disasters as
  Interactive Components of Global Ecodynamics}}}
  (\bibinfo{publisher}{Springer}, \bibinfo{address}{Berlin},
  \bibinfo{year}{2006}).

\bibitem[{\citenamefont{Bogachev
  et~al.}(2008{\natexlab{a}})\citenamefont{Bogachev, Eichner, and
  Bunde}}]{Bogachev-Eichner-Bunde-PAG-2008}
\bibinfo{author}{\bibfnamefont{M.~I.} \bibnamefont{Bogachev}},
  \bibinfo{author}{\bibfnamefont{J.~F.} \bibnamefont{Eichner}},
  \bibnamefont{and} \bibinfo{author}{\bibfnamefont{A.}~\bibnamefont{Bunde}},
  \bibinfo{journal}{Pure Appl. Geophys.} \textbf{\bibinfo{volume}{165}},
  \bibinfo{pages}{1195} (\bibinfo{year}{2008}{\natexlab{a}}).

\bibitem[{\citenamefont{Bunde et~al.}(2003)\citenamefont{Bunde, Eichner,
  Havlin, and Kantelhardt}}]{Bunde-Eichner-Havlin-Kantelhardt-2003-PA}
\bibinfo{author}{\bibfnamefont{A.}~\bibnamefont{Bunde}},
  \bibinfo{author}{\bibfnamefont{J.~F.} \bibnamefont{Eichner}},
  \bibinfo{author}{\bibfnamefont{S.}~\bibnamefont{Havlin}}, \bibnamefont{and}
  \bibinfo{author}{\bibfnamefont{J.~W.} \bibnamefont{Kantelhardt}},
  \bibinfo{journal}{Physica A} \textbf{\bibinfo{volume}{330}},
  \bibinfo{pages}{1} (\bibinfo{year}{2003}).

\bibitem[{\citenamefont{Bunde et~al.}(2005{\natexlab{b}})\citenamefont{Bunde,
  Eichner, Kantelhardt, and
  Havlin}}]{Bunde-Eichner-Kantelhardt-Havlin-2005-PRL}
\bibinfo{author}{\bibfnamefont{A.}~\bibnamefont{Bunde}},
  \bibinfo{author}{\bibfnamefont{J.~F.} \bibnamefont{Eichner}},
  \bibinfo{author}{\bibfnamefont{J.~W.} \bibnamefont{Kantelhardt}},
  \bibnamefont{and} \bibinfo{author}{\bibfnamefont{S.}~\bibnamefont{Havlin}},
  \bibinfo{journal}{Phys. Rev. Lett.} \textbf{\bibinfo{volume}{94}},
  \bibinfo{pages}{048701} (\bibinfo{year}{2005}{\natexlab{b}}).

\bibitem[{\citenamefont{Altmann and Kantz}(2005)}]{Altmann-Kantz-2005-PRE}
\bibinfo{author}{\bibfnamefont{E.~G.} \bibnamefont{Altmann}} \bibnamefont{and}
  \bibinfo{author}{\bibfnamefont{H.}~\bibnamefont{Kantz}},
  \bibinfo{journal}{Phys. Rev. E} \textbf{\bibinfo{volume}{71}},
  \bibinfo{pages}{056106} (\bibinfo{year}{2005}).

\bibitem[{\citenamefont{Olla}(2007)}]{Olla-2007-PRE}
\bibinfo{author}{\bibfnamefont{P.}~\bibnamefont{Olla}}, \bibinfo{journal}{Phys.
  Rev. E} \textbf{\bibinfo{volume}{76}}, \bibinfo{pages}{011122}
  (\bibinfo{year}{2007}).

\bibitem[{\citenamefont{Santhanam and Kantz}(2008)}]{Santhanam-Kantz-2008-PRE}
\bibinfo{author}{\bibfnamefont{M.~S.} \bibnamefont{Santhanam}}
  \bibnamefont{and} \bibinfo{author}{\bibfnamefont{H.}~\bibnamefont{Kantz}},
  \bibinfo{journal}{Phys. Rev. E} \textbf{\bibinfo{volume}{78}},
  \bibinfo{pages}{051113} (\bibinfo{year}{2008}).

\bibitem[{\citenamefont{Eichner et~al.}(2007)\citenamefont{Eichner,
  Kantelhardt, Bunde, and Havlin}}]{Eichner-Kantelhardt-Bunde-Havlin-2007-PRE}
\bibinfo{author}{\bibfnamefont{J.~F.} \bibnamefont{Eichner}},
  \bibinfo{author}{\bibfnamefont{J.~W.} \bibnamefont{Kantelhardt}},
  \bibinfo{author}{\bibfnamefont{A.}~\bibnamefont{Bunde}}, \bibnamefont{and}
  \bibinfo{author}{\bibfnamefont{S.}~\bibnamefont{Havlin}},
  \bibinfo{journal}{Phys. Rev. E} \textbf{\bibinfo{volume}{75}},
  \bibinfo{pages}{011128} (\bibinfo{year}{2007}).

\bibitem[{\citenamefont{Meneveau and
  Sreenivasan}(1987)}]{Meneveau-Sreenivasan-1987-PRL}
\bibinfo{author}{\bibfnamefont{C.}~\bibnamefont{Meneveau}} \bibnamefont{and}
  \bibinfo{author}{\bibfnamefont{K.~R.} \bibnamefont{Sreenivasan}},
  \bibinfo{journal}{Phys. Rev. Lett.} \textbf{\bibinfo{volume}{59}},
  \bibinfo{pages}{1424} (\bibinfo{year}{1987}).

\bibitem[{\citenamefont{Bacry et~al.}(2001{\natexlab{a}})\citenamefont{Bacry,
  Delour, and Muzy}}]{Bacry-Delour-Muzy-2001-PRE}
\bibinfo{author}{\bibfnamefont{E.}~\bibnamefont{Bacry}},
  \bibinfo{author}{\bibfnamefont{J.}~\bibnamefont{Delour}}, \bibnamefont{and}
  \bibinfo{author}{\bibfnamefont{J.-F.} \bibnamefont{Muzy}},
  \bibinfo{journal}{Phys. Rev. E} \textbf{\bibinfo{volume}{64}},
  \bibinfo{pages}{026103} (\bibinfo{year}{2001}{\natexlab{a}}).

\bibitem[{\citenamefont{Bogachev et~al.}(2007)\citenamefont{Bogachev, Eichner,
  and Bunde}}]{Bogachev-Eichner-Bunde-2007-PRL}
\bibinfo{author}{\bibfnamefont{M.~I.} \bibnamefont{Bogachev}},
  \bibinfo{author}{\bibfnamefont{J.~F.} \bibnamefont{Eichner}},
  \bibnamefont{and} \bibinfo{author}{\bibfnamefont{A.}~\bibnamefont{Bunde}},
  \bibinfo{journal}{Phys. Rev. Lett.} \textbf{\bibinfo{volume}{99}},
  \bibinfo{pages}{240601} (\bibinfo{year}{2007}).

\bibitem[{\citenamefont{Bogachev
  et~al.}(2008{\natexlab{b}})\citenamefont{Bogachev, Eichner, and
  Bunde}}]{Bogachev-Eichner-Bunde-EPJST-2008}
\bibinfo{author}{\bibfnamefont{M.~I.} \bibnamefont{Bogachev}},
  \bibinfo{author}{\bibfnamefont{J.~F.} \bibnamefont{Eichner}},
  \bibnamefont{and} \bibinfo{author}{\bibfnamefont{A.}~\bibnamefont{Bunde}},
  \bibinfo{journal}{Eur. Phys. J. Spec. Top.} \textbf{\bibinfo{volume}{161}},
  \bibinfo{pages}{181} (\bibinfo{year}{2008}{\natexlab{b}}).

\bibitem[{\citenamefont{Bogachev and Bunde}(2008)}]{Bogachev-Bunde-2008-PRE}
\bibinfo{author}{\bibfnamefont{M.~I.} \bibnamefont{Bogachev}} \bibnamefont{and}
  \bibinfo{author}{\bibfnamefont{A.}~\bibnamefont{Bunde}},
  \bibinfo{journal}{Phys. Rev. E} \textbf{\bibinfo{volume}{78}},
  \bibinfo{pages}{036114} (\bibinfo{year}{2008}).

\bibitem[{\citenamefont{Muzy et~al.}(2000)\citenamefont{Muzy, Delour, and
  Bacry}}]{Muzy-Delour-Bacry-2000-EPJB}
\bibinfo{author}{\bibfnamefont{J.-F.} \bibnamefont{Muzy}},
  \bibinfo{author}{\bibfnamefont{J.}~\bibnamefont{Delour}}, \bibnamefont{and}
  \bibinfo{author}{\bibfnamefont{E.}~\bibnamefont{Bacry}},
  \bibinfo{journal}{Eur. Phys. J. B} \textbf{\bibinfo{volume}{17}},
  \bibinfo{pages}{537} (\bibinfo{year}{2000}).

\bibitem[{\citenamefont{Bacry et~al.}(2001{\natexlab{b}})\citenamefont{Bacry,
  Delour, and Muzy}}]{Bacry-Delour-Muzy-2001-PA}
\bibinfo{author}{\bibfnamefont{E.}~\bibnamefont{Bacry}},
  \bibinfo{author}{\bibfnamefont{J.}~\bibnamefont{Delour}}, \bibnamefont{and}
  \bibinfo{author}{\bibfnamefont{J.-F.} \bibnamefont{Muzy}},
  \bibinfo{journal}{Physica A} \textbf{\bibinfo{volume}{299}},
  \bibinfo{pages}{84} (\bibinfo{year}{2001}{\natexlab{b}}).

\bibitem[{\citenamefont{Yamasaki et~al.}(2005)\citenamefont{Yamasaki, Muchnik,
  Havlin, Bunde, and
  Stanley}}]{Yamasaki-Muchnik-Havlin-Bunde-Stanley-2005-PNAS}
\bibinfo{author}{\bibfnamefont{K.}~\bibnamefont{Yamasaki}},
  \bibinfo{author}{\bibfnamefont{L.}~\bibnamefont{Muchnik}},
  \bibinfo{author}{\bibfnamefont{S.}~\bibnamefont{Havlin}},
  \bibinfo{author}{\bibfnamefont{A.}~\bibnamefont{Bunde}}, \bibnamefont{and}
  \bibinfo{author}{\bibfnamefont{H.~E.} \bibnamefont{Stanley}},
  \bibinfo{journal}{Proc. Natl. Acad. Sci. USA} \textbf{\bibinfo{volume}{102}},
  \bibinfo{pages}{9424} (\bibinfo{year}{2005}).

\bibitem[{\citenamefont{Wang et~al.}(2006)\citenamefont{Wang, Yamasaki, Havlin,
  and Stanley}}]{Wang-Yamasaki-Havlin-Stanley-2006-PRE}
\bibinfo{author}{\bibfnamefont{F.-Z.} \bibnamefont{Wang}},
  \bibinfo{author}{\bibfnamefont{K.}~\bibnamefont{Yamasaki}},
  \bibinfo{author}{\bibfnamefont{S.}~\bibnamefont{Havlin}}, \bibnamefont{and}
  \bibinfo{author}{\bibfnamefont{H.~E.} \bibnamefont{Stanley}},
  \bibinfo{journal}{Phys. Rev. E} \textbf{\bibinfo{volume}{73}},
  \bibinfo{pages}{026117} (\bibinfo{year}{2006}).

\bibitem[{\citenamefont{Wang et~al.}(2007)\citenamefont{Wang, Weber, Yamasaki,
  Havlin, and Stanley}}]{Wang-Weber-Yamasaki-Havlin-Stanley-2007-EPJB}
\bibinfo{author}{\bibfnamefont{F.}~\bibnamefont{Wang}},
  \bibinfo{author}{\bibfnamefont{P.}~\bibnamefont{Weber}},
  \bibinfo{author}{\bibfnamefont{K.}~\bibnamefont{Yamasaki}},
  \bibinfo{author}{\bibfnamefont{S.}~\bibnamefont{Havlin}}, \bibnamefont{and}
  \bibinfo{author}{\bibfnamefont{H.~E.} \bibnamefont{Stanley}},
  \bibinfo{journal}{Eur. Phys. J. B} \textbf{\bibinfo{volume}{55}},
  \bibinfo{pages}{123} (\bibinfo{year}{2007}).

\bibitem[{\citenamefont{Jung et~al.}(2008)\citenamefont{Jung, Wang, Havlin,
  Kaizoji, Moon, and
  Stanley}}]{Jung-Wang-Havlin-Kaizoji-Moon-Stanley-2008-EPJB}
\bibinfo{author}{\bibfnamefont{W.-S.} \bibnamefont{Jung}},
  \bibinfo{author}{\bibfnamefont{F.-Z.} \bibnamefont{Wang}},
  \bibinfo{author}{\bibfnamefont{S.}~\bibnamefont{Havlin}},
  \bibinfo{author}{\bibfnamefont{T.}~\bibnamefont{Kaizoji}},
  \bibinfo{author}{\bibfnamefont{H.~T.} \bibnamefont{Moon}}, \bibnamefont{and}
  \bibinfo{author}{\bibfnamefont{H.~E.} \bibnamefont{Stanley}},
  \bibinfo{journal}{Eur. Phys. J. B} \textbf{\bibinfo{volume}{62}},
  \bibinfo{pages}{113} (\bibinfo{year}{2008}).

\bibitem[{\citenamefont{Qiu et~al.}(2008)\citenamefont{Qiu, Guo, and
  Chen}}]{Qiu-Guo-Chen-2008-PA}
\bibinfo{author}{\bibfnamefont{T.}~\bibnamefont{Qiu}},
  \bibinfo{author}{\bibfnamefont{L.}~\bibnamefont{Guo}}, \bibnamefont{and}
  \bibinfo{author}{\bibfnamefont{G.}~\bibnamefont{Chen}},
  \bibinfo{journal}{Physica A} \textbf{\bibinfo{volume}{387}},
  \bibinfo{pages}{6812} (\bibinfo{year}{2008}).

\bibitem[{\citenamefont{Wang et~al.}(2008)\citenamefont{Wang, Yamasaki, Havlin,
  and Stanley}}]{Wang-Yamasaki-Havlin-Stanley-2008-PRE}
\bibinfo{author}{\bibfnamefont{F.-Z.} \bibnamefont{Wang}},
  \bibinfo{author}{\bibfnamefont{K.}~\bibnamefont{Yamasaki}},
  \bibinfo{author}{\bibfnamefont{S.}~\bibnamefont{Havlin}}, \bibnamefont{and}
  \bibinfo{author}{\bibfnamefont{H.~E.} \bibnamefont{Stanley}},
  \bibinfo{journal}{Phys. Rev. E} \textbf{\bibinfo{volume}{77}},
  \bibinfo{pages}{016109} (\bibinfo{year}{2008}).

\bibitem[{\citenamefont{Lee et~al.}(2006)\citenamefont{Lee, Lee, and
  Rikvold}}]{Lee-Lee-Rikvold-2006-JKPS}
\bibinfo{author}{\bibfnamefont{J.~W.} \bibnamefont{Lee}},
  \bibinfo{author}{\bibfnamefont{K.~E.} \bibnamefont{Lee}}, \bibnamefont{and}
  \bibinfo{author}{\bibfnamefont{P.~A.} \bibnamefont{Rikvold}},
  \bibinfo{journal}{J. Korean Phys. Soc.} \textbf{\bibinfo{volume}{48}},
  \bibinfo{pages}{S123} (\bibinfo{year}{2006}).

\bibitem[{\citenamefont{Ren and Zhou}(2008)}]{Ren-Zhou-2008-EPL}
\bibinfo{author}{\bibfnamefont{F.}~\bibnamefont{Ren}} \bibnamefont{and}
  \bibinfo{author}{\bibfnamefont{W.-X.} \bibnamefont{Zhou}},
  \bibinfo{journal}{EPL} \textbf{\bibinfo{volume}{84}}, \bibinfo{pages}{68001}
  (\bibinfo{year}{2008}).

\bibitem[{\citenamefont{Ren et~al.}(2009)\citenamefont{Ren, Guo, and
  Zhou}}]{Ren-Guo-Zhou-2009-PA}
\bibinfo{author}{\bibfnamefont{F.}~\bibnamefont{Ren}},
  \bibinfo{author}{\bibfnamefont{L.}~\bibnamefont{Guo}}, \bibnamefont{and}
  \bibinfo{author}{\bibfnamefont{W.-X.} \bibnamefont{Zhou}},
  \bibinfo{journal}{Physica A} \textbf{\bibinfo{volume}{388}},
  \bibinfo{pages}{XXX} (\bibinfo{year}{2009}).

\bibitem[{\citenamefont{Anselmet et~al.}(1984)\citenamefont{Anselmet, Gagne,
  Hopfinger, and Antonia}}]{Anselmet-Gagne-Hopfinger-Antonia-1984-JFM}
\bibinfo{author}{\bibfnamefont{F.}~\bibnamefont{Anselmet}},
  \bibinfo{author}{\bibfnamefont{Y.}~\bibnamefont{Gagne}},
  \bibinfo{author}{\bibfnamefont{E.~J.} \bibnamefont{Hopfinger}},
  \bibnamefont{and} \bibinfo{author}{\bibfnamefont{R.~A.}
  \bibnamefont{Antonia}}, \bibinfo{journal}{J. Fluid Mech.}
  \textbf{\bibinfo{volume}{140}}, \bibinfo{pages}{63} (\bibinfo{year}{1984}).

\bibitem[{\citenamefont{Zhou and Sornette}(2002)}]{Zhou-Sornette-2002-PD}
\bibinfo{author}{\bibfnamefont{W.-X.} \bibnamefont{Zhou}} \bibnamefont{and}
  \bibinfo{author}{\bibfnamefont{D.}~\bibnamefont{Sornette}},
  \bibinfo{journal}{Physica D} \textbf{\bibinfo{volume}{165}},
  \bibinfo{pages}{94} (\bibinfo{year}{2002}).

\bibitem[{\citenamefont{Peng et~al.}(1994)\citenamefont{Peng, Buldyrev, Havlin,
  Simons, Stanley, and
  Goldberger}}]{Peng-Buldyrev-Havlin-Simons-Stanley-Goldberger-1994-PRE}
\bibinfo{author}{\bibfnamefont{C.-K.} \bibnamefont{Peng}},
  \bibinfo{author}{\bibfnamefont{S.~V.} \bibnamefont{Buldyrev}},
  \bibinfo{author}{\bibfnamefont{S.}~\bibnamefont{Havlin}},
  \bibinfo{author}{\bibfnamefont{M.}~\bibnamefont{Simons}},
  \bibinfo{author}{\bibfnamefont{H.~E.} \bibnamefont{Stanley}},
  \bibnamefont{and} \bibinfo{author}{\bibfnamefont{A.~L.}
  \bibnamefont{Goldberger}}, \bibinfo{journal}{Phys. Rev. E}
  \textbf{\bibinfo{volume}{49}}, \bibinfo{pages}{1685} (\bibinfo{year}{1994}).

\bibitem[{\citenamefont{Kantelhardt et~al.}(2001)\citenamefont{Kantelhardt,
  Koscielny-Bunde, Rego, Havlin, and
  Bunde}}]{Kantelhardt-Bunde-Rego-Havlin-Bunde-2001-PA}
\bibinfo{author}{\bibfnamefont{J.~W.} \bibnamefont{Kantelhardt}},
  \bibinfo{author}{\bibfnamefont{E.}~\bibnamefont{Koscielny-Bunde}},
  \bibinfo{author}{\bibfnamefont{H.~H.~A.} \bibnamefont{Rego}},
  \bibinfo{author}{\bibfnamefont{S.}~\bibnamefont{Havlin}}, \bibnamefont{and}
  \bibinfo{author}{\bibfnamefont{A.}~\bibnamefont{Bunde}},
  \bibinfo{journal}{Physica A} \textbf{\bibinfo{volume}{295}},
  \bibinfo{pages}{441} (\bibinfo{year}{2001}).

\bibitem[{\citenamefont{Kantelhardt et~al.}(2002)\citenamefont{Kantelhardt,
  Zschiegner, Koscielny-Bunde, Havlin, Bunde, and
  Stanley}}]{Kantelhardt-Zschiegner-Bunde-Havlin-Bunde-Stanley-2002-PA}
\bibinfo{author}{\bibfnamefont{J.~W.} \bibnamefont{Kantelhardt}},
  \bibinfo{author}{\bibfnamefont{S.~A.} \bibnamefont{Zschiegner}},
  \bibinfo{author}{\bibfnamefont{E.}~\bibnamefont{Koscielny-Bunde}},
  \bibinfo{author}{\bibfnamefont{S.}~\bibnamefont{Havlin}},
  \bibinfo{author}{\bibfnamefont{A.}~\bibnamefont{Bunde}}, \bibnamefont{and}
  \bibinfo{author}{\bibfnamefont{H.~E.} \bibnamefont{Stanley}},
  \bibinfo{journal}{Physica A} \textbf{\bibinfo{volume}{316}},
  \bibinfo{pages}{87} (\bibinfo{year}{2002}).

\end{thebibliography}

\end{document}